\documentclass[12pt,a4paper]{article}
\usepackage{graphicx}
\usepackage{times}
\usepackage{natbib} 
\usepackage{float}
\usepackage{xcolor}
\usepackage{hyperref}
\hypersetup{
    colorlinks=true,
    linkcolor=black,
    filecolor=magenta,      
    urlcolor=blue,
}
\title{\bf Study of pulsational instabilities in models of $\delta$\,Scuti pulsators discovered under the Nainital-Cape Survey}%\footnote{A potential footnote to the title may be added here}}

\author{Abhay Pratap Yadav\thanks{E-mail: abhaypratapbhu@yahoo.com} \\
\vspace{0.5cm}\\
\normalsize  Government Model College Shahpura, Dindori 481990, India\\ 
}
\date{\mbox{}}
\begin{document}
\maketitle
\setcounter{page}{1}
\pagestyle{plain}
    \makeatletter
    \renewcommand*{\pagenumbering}[1]{%
       \gdef\thepage{\csname @#1\endcsname\c@page}%
    }
    \makeatother
\pagenumbering{arabic}

%
% WE REDEFINE THE plain LaTeX PAGESTYLE !!! 
% THIS PAGESTYLE WILL BE USED FOR THE FIRST PAGE ONLY !
% Please do not change the following lines
%
\def\bull{\vrule height .9ex width .8ex depth -.1ex}
\makeatletter
\def\ps@plain{\let\@mkboth\gobbletwo
\def\@oddhead{}\def\@oddfoot{\hfil\scriptsize\bull\quad
"2nd Belgo-Indian Network for Astronomy \& astrophysics (BINA) workshop'', held in Brussels (Belgium), 9-12 October 2018 \quad\bull}%
\def\@evenhead{}\let\@evenfoot\@oddfoot}
\makeatother
%
% AND DEFINE OUR MACROS FOR THE REFERENCE LIST
% I.E \beginrefer \refer and \endrefer
%
\def\beginrefer{\section*{References}%
\begin{quotation}\mbox{}\par}
\def\refer#1\par{{\setlength{\parindent}{-\leftmargin}\indent#1\par}}
\def\endrefer{\end{quotation}}
%
% BEGIN THE ABSTRACT WITH \noindent\small, ENCLOSE IT IN A GROUP
% AND BOLDFACE THE TITLE.
%
{\noindent\small{\bf Abstract:} 
Under the Nainital-Cape Survey, eight $\delta$\,Scuti type pulsators have been discovered with 
the pulsation periods in the range of several minutes to few hours. In order to understand these
observed pulsational variabilities, we have  performed non-adiabatic linear stability analyses in models
of these stars having mass in the range of 1 to 3 M$_{\odot}$. Several low order p-modes are found to be unstable where the pulsation periods
associated with these unstable modes are in good agreement with the observed periods. Particularly for HD\,118660,  HD\,113878, HD\,102480, 
HD\,98851, and HD\,25515, 
we demonstrate that the observed variabilities can be explained with the low order radial p-mode pulsations.
}
\vspace{0.5cm}\\
% SPECIFY UP TO 5 KEYWORDS SEPARATED BY ' -- '
\section{Introduction}

The Nainital-Cape survey started in the late 1990s with the aim to 
search and study pulsations in Ap and Am stars using high-speed photometry and high-resolution spectroscopy. 
Under this survey, more than 300 stars are monitored. This resulted in the discovery of 
eight new $\delta$\,Scuti ($\delta$\,Sct) type pulsators 
(for the detailed results of the survey, see e.g., Ashoka et al. 2000; Balona et al. 2013, 2016; Martinez et al. 2001; Joshi et al. 2003,
2006, 2009, 2010, 2012a, 2012b, 2014, 2016). 
The $\delta$\,Sct pulsators are main sequence Pop $\rm{I}$ stars having mass in the range between 1.5 and 2.5 M$_{\odot}$. 
Pulsation periods of these stars lie in the range between 18 minutes to 8 hours (Aerts et al. 2010). These pulsations 
are believed to be driven by the $\kappa$-mechanism operating in the second ionization zone of helium. However, detailed asteroseismic
modelling of these stars is required to understand the excitation mechanism and pulsation properties. So far, 
asteroseismic modelling of the $\delta$\,Sct pulsators discovered under the Nainital-Cape survey has not been done. 
This motivated us 
 to perform the modelling for these eight stars to enhance our present understanding of their pulsational
variabilities. 
Most of the atmospheric parameters, including chemical 
compositions, for these eight $\delta$\,Sct pulsators have been compiled by Joshi et al. (2017) using observed and synthetic spectra.
 In this survey, several monitored stars do not exhibit any photometric variabilities to the detection limit of the observations used. 
 Hence it would be interesting to 
find the cause for the absence of pulsations in this set of observed stars.

Out of eight pulsators discovered under the Nainital-Cape Survey, the present study is primarily focused on 
HD\,13038, HD\,13079, HD\,25515, HD\,98851, HD\,102480, HD\,113878, and HD\,118660 (Table\,1).
Since the star HD\,12098 is a part of another ongoing asteroseismic analysis, it has been excluded in the present study. 
HD\,12098 is a rapid oscillator with a period of 7 minutes, classified as a F0-type star (Martinez et al. 2001).

The models are described in Section\,\ref{a}. The method used for the linear stability analysis is outlined in Section\,\ref{b},
followed by the results in Section\,\ref{c}. We end with a discussion and our conclusions in Section\,\ref{d}.

\section{Stellar Models}
\label{a}
\begin{figure}[h]
\centering $
\large
\begin{array}{c}
\scalebox{0.76}{ \input{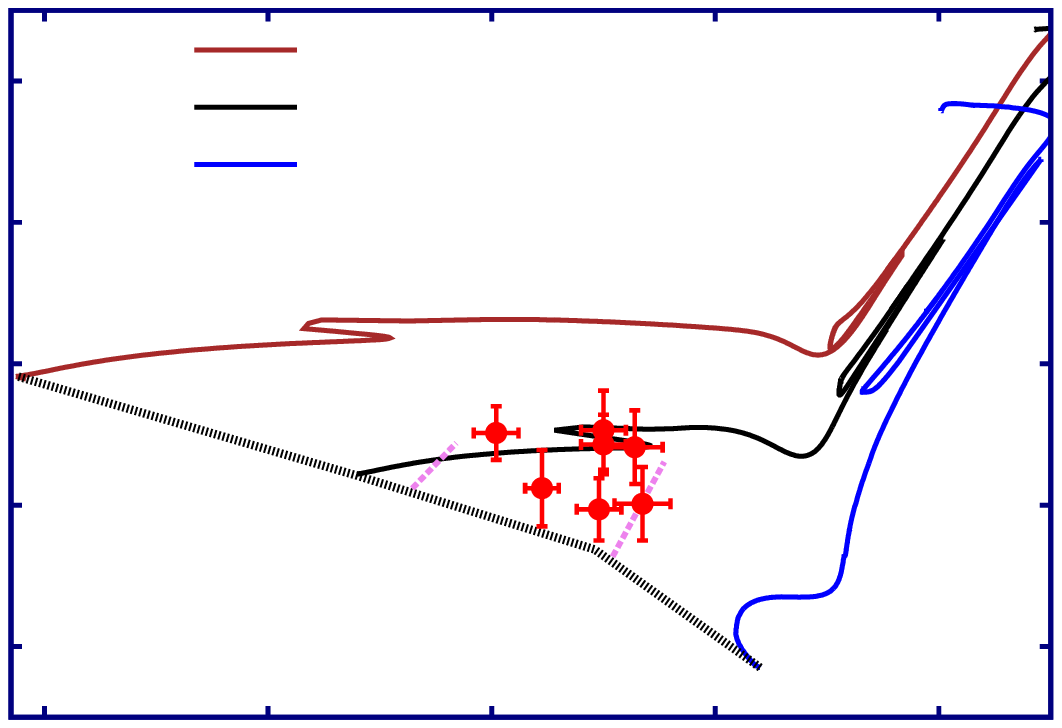} } 
 \end{array}$
 \caption{Evolutionary tracks of models having a mass 1\,M$_{\odot}$ (blue line), 2\,M$_{\odot}$ (black line), and 3\,M$_{\odot}$ (brown line) 
 are given from the zero-age main sequence (grey line). The red dots with error bars represent 
 the location of the 7 considered stars while the violet lines denote the blue (left) and red (right) edge of the theoretical instability strip of 
$\delta$\,Sct stars for low order radial modes as calculated by Dupret et al. (2005). }
 \normalsize
 \label{hrd}
 \end{figure}

In order to study the instabilities present in the $\delta$\,Sct stars, we have considered stellar models having mass ($M$) in the
range of 
1 and 3 M$_{\odot}$ with solar chemical composition (X = 0.70, Y = 0.28, and Z = 0.02, where the fraction of hydrogen, helium, and heavier
elements are represented by X, Y and Z, respectively). The effective temperature ($T_{\rm eff}$) and luminosity ($\log L/L_{\odot}$) of 
these seven stars 
as given in Table\,1 are taken from Joshi et al. (2017). Fig.\,\ref{hrd} shows the locations of these 
stars in the Hertzsprung-Russell 
diagram (HRD) along with the evolutionary
tracks of the stellar models having mass of 1, 2 and 3 M$_{\odot}$ with solar chemical composition. These evolutionary 
tracks have been generated using the `mad star EZ' code\footnote{\url{http://www.astro.wisc.edu/~townsend/static.php?ref=ez-web}}. The 
position of the theoretical $\delta$\,Sct instability strip for low order radial modes is also 
given on Fig.\,\ref{hrd} (cf. Dupret et al. 2004, 2005). The location of the considered stars is compatible with
the theoretical 
instability strip of $\delta$\,Sct stars. Therefore, these stars are likely to have pulsation periods in the range of 18 minutes to 8 hours 
(see e.g., Aerts et al. 2010) as found in several other $\delta$\,Sct pulsators (see e.g., Breger et al. 2005). 
\begin{table}[h]
\caption{$T_{\rm eff}$ and $\log L/L_{\odot}$ of the considered stars (values published by Joshi et al. 2017).} 
\small
\begin{center} 
\begin{tabular}{| c | c c |}
\hline 
Star & $T_{\rm eff}$ (K) & $\log L/L_{\odot}$ \\
\hline
HD\,13038 & 7960 $\pm$ 200 & 1.51 $\pm$ 0.19\\
HD\,13079 & 7040 $\pm$ 200 & 0.97 $\pm$ 0.22\\
HD\,25515 & 6650 $\pm$ 250 & 1.01 $\pm$ 0.26\\
HD\,98851 & 7000 $\pm$ 200 & 1.43 $\pm$ 0.21 \\
HD\,102480 & 6720 $\pm$ 250 & 1.41 $\pm$ 0.26\\
HD\,113878 & 7000 $\pm$ 200 & 1.53 $\pm$ 0.28 \\
HD\,118660 & 7550 $\pm$ 150 & 1.12 $\pm$ 0.27 \\

\hline 
\end{tabular} 
\end{center} 
\end{table} 

Using the values of $M$, $T_{\rm eff}$ and $\log L/L_{\odot}$, stellar envelope models have been constructed by integrating the equation of mass 
conservation, hydrostatic equilibrium and energy transport from the photosphere up to a chosen maximum cut-off temperature. The value of 
the cut-off 
temperature 
(of the order of 10$^{7}$ K) has been 
taken in such a way that almost the entire part of the stellar envelopes can be investigated for instabilities. Since the expected instabilities for   
$\delta$\,Sct stars operate in the ionization zones, the relevant part of the stars has been fully considered. Stefan-Boltzmann's law and the 
photospheric pressure (as given by Kippenhahn et al. 2012) have been taken as initial boundary conditions for the integration. For the opacities, 
OPAL tables  (Rogers \& Iglesias 1992; Rogers et al. 1996; Iglesias \& Rogers 1996) have been used. The onset of convection is prescribed by
the Schwarzschild's criterion. The standard mixing length theory 
(B\"ohm-Vitense 1958) is used for the convection with a value of 1.5 pressure scale heights for the mixing length parameter $\alpha$.  
Dupret et al. (2004) have shown that the excitation of modes is influenced by the treatment of convection as well as the choice of $\alpha$.
In order to examine the effect of $\alpha$ on the instabilities, we have used two additional sets of 
stellar models of the star HD\,13038 with $\alpha$\,= 1.8 and 2.0, respectively.

\section{Linear stability analysis}
\label{b}
 The present study is restricted to the radial perturbations (pulsation modes with degree $\ell = 0$) where the equations governing the 
 stellar stability and pulsations are taken 
 in the form as given by Gautschy \& Glatzel (1990b). These equations, together with boundary conditions, form a fourth order eigenvalue 
 problem which is solved using the Riccati method as mentioned by Gautschy \& Glatzel (1990a). 
 The resulting eigenfrequencies are complex ($\sigma_{\rm{r}}$ + i$\sigma_{\rm{i}}$) where the real part ($\sigma_{\rm{r}}$) 
 is associated with the pulsation periods 
 and the imaginary part ($\sigma_{\rm{i}}$) provides information about damping ($\sigma_{\rm{i}}$ $>$ 0) or excitation ($\sigma_{\rm{i}}$ $<$ 0) 
 of the considered mode. The eigenfrequencies are normalized with the global free fall time $\sqrt{R^3/3\,GM}$ where $R$, $G$ and $M$ 
 are the stellar radius, the gravitational constant and the stellar mass, respectively. For the treatment of convection, we have used `frozen in approximation' as 
 introduced by Baker \& Kippenhahn (1965). Under this approximation, the Lagrangian perturbation of the convective flux is disregarded.

\section{Results}
\label{c}
\begin{figure*}[h]
\centering $
\large
\begin{array}{cc}

   \scalebox{0.66}{ \input{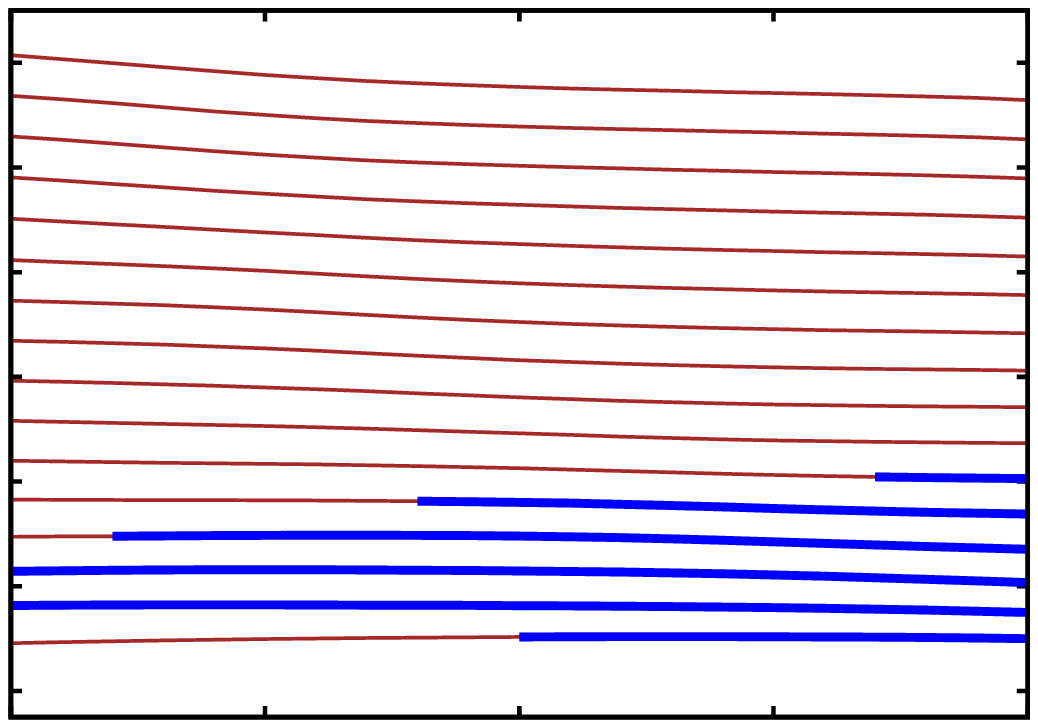} } 
    \scalebox{0.66}{ \input{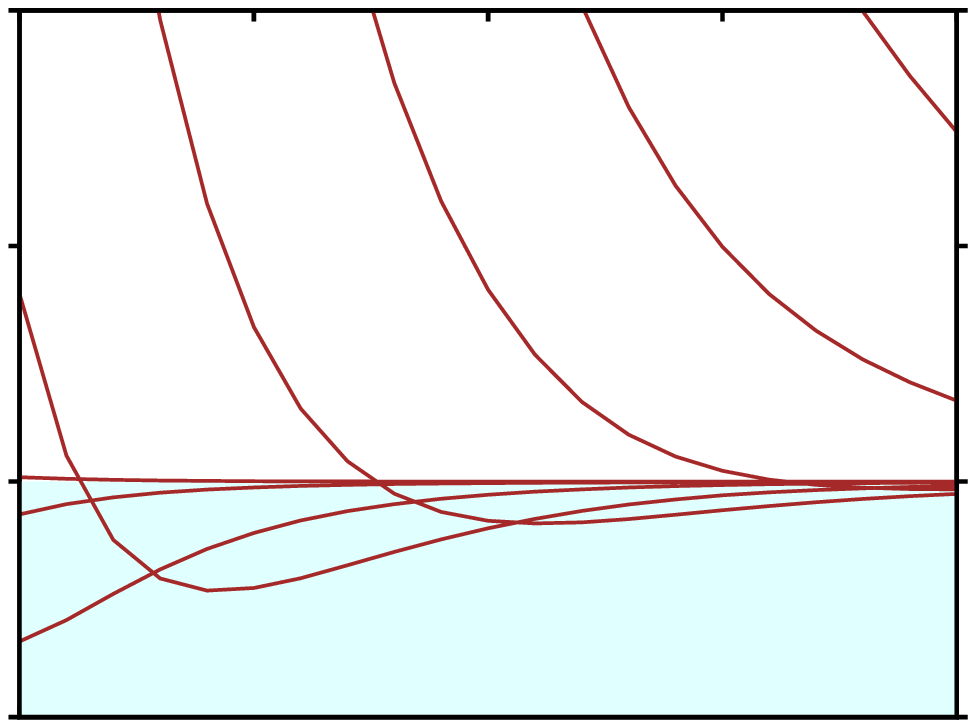} } 
 \end{array}$
 \caption{Modal diagram for HD\,113878. The real (a) and imaginary (b) parts of the eigenfrequencies of radial modes ($\ell$\,=\,0)
 normalized with the global free fall timescales are given 
 as a function of stellar mass for the models of HD\,113878, where $T_{\rm eff}$ = 7000 K and $\log L/L_{\odot}$ = 1.53. 
 Real parts represented by a thick line (a) and 
 negative values of the imaginary part in (b) correspond to excited modes. The region with negative values for the imaginary part 
 has a light blue background in (b).}
 \normalsize
 \label{m_hd113878}
 \end{figure*}

Linear stability analyses have been performed in the models of
seven $\delta$\,Sct stars discovered under the Nainital-Cape survey. 
The representation of eigenfrequencies as a function of a stellar 
parameter, particularly $M$ or $T_{\rm eff}$, is known as a `Modal diagram' (for more details, see, e.g., Saio et al. 1998).
The real ($\sigma_{\rm{r}}$) and imaginary ($\sigma_{\rm{i}}$) 
parts of the eigenfrequency are given as a function of stellar mass for the star HD\,113878 in Fig.\,\ref{m_hd113878}.  
Blue thick lines in Fig.\,\ref{m_hd113878}(a) and negative
value of the imaginary parts in Fig.\,\ref{m_hd113878}(b) represent the unstable modes. In Fig.\,\ref{m_hd113878}(b), 
the region with $\sigma_{\rm{i}}$ $<$ 0 is filled with light blue.
For this star, low order modes are excited. Instabilities associated with these modes are weak compared to their
counterparts in more massive stars of the upper main sequence (see, e.g., Yadav \& Glatzel 2017a) which indicates
that the $\delta$\,Sct stars are likely to have smaller amplitude pulsations compared to
more massive stars. Modes in Fig.\,\ref{m_hd113878}(a) are well-spaced and
mode coupling phenomena, as observed in the models of massive and evolved stars 
(see, e.g., Glatzel \& Kiriakidis 1993; Glatzel et al. 1993; Kiriakidis et al. 1993; Yadav \& Glatzel 2017b), are absent. 
Fig.\,\ref{m_hd113878}(a) also indicates
that the high order modes are completely damped and that the number of unstable modes increases as a function of stellar mass. 
In case of HD\,113878, we have only two unstable modes for the model with mass 1 M$_{\odot}$ while six modes are unstable for the 
model having a mass of 3 M$_{\odot}$ (Fig.\,\ref{m_hd113878}).

\begin{figure*}
\centering $
\large
\begin{array}{cc}

   \scalebox{0.66}{ \input{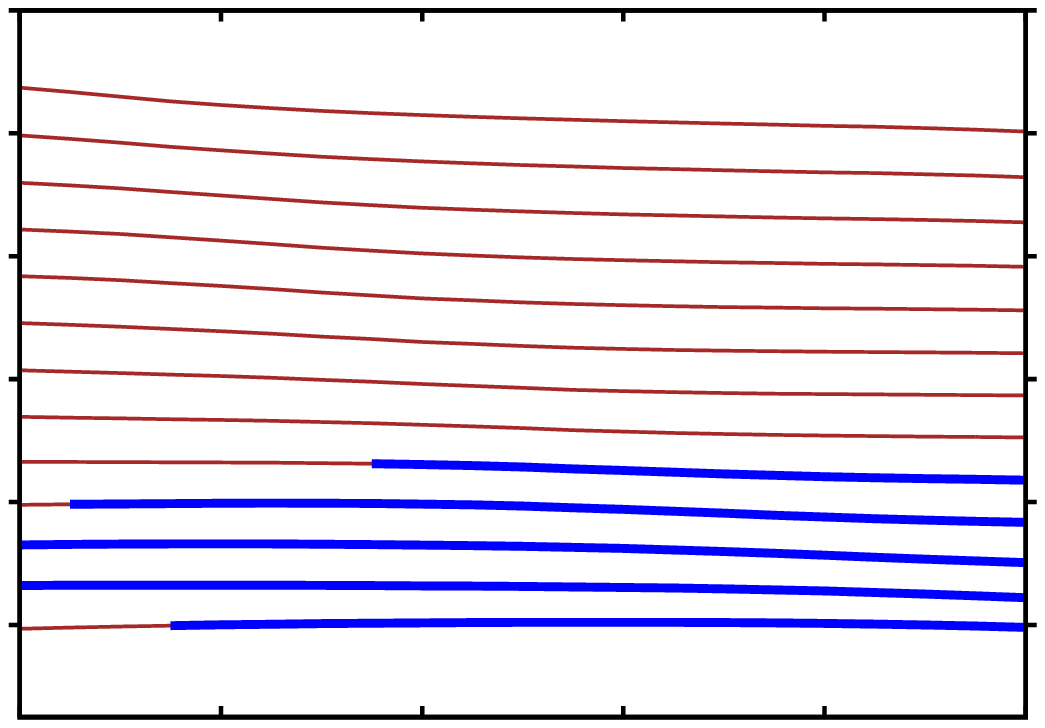} } 
    \scalebox{0.66}{ \input{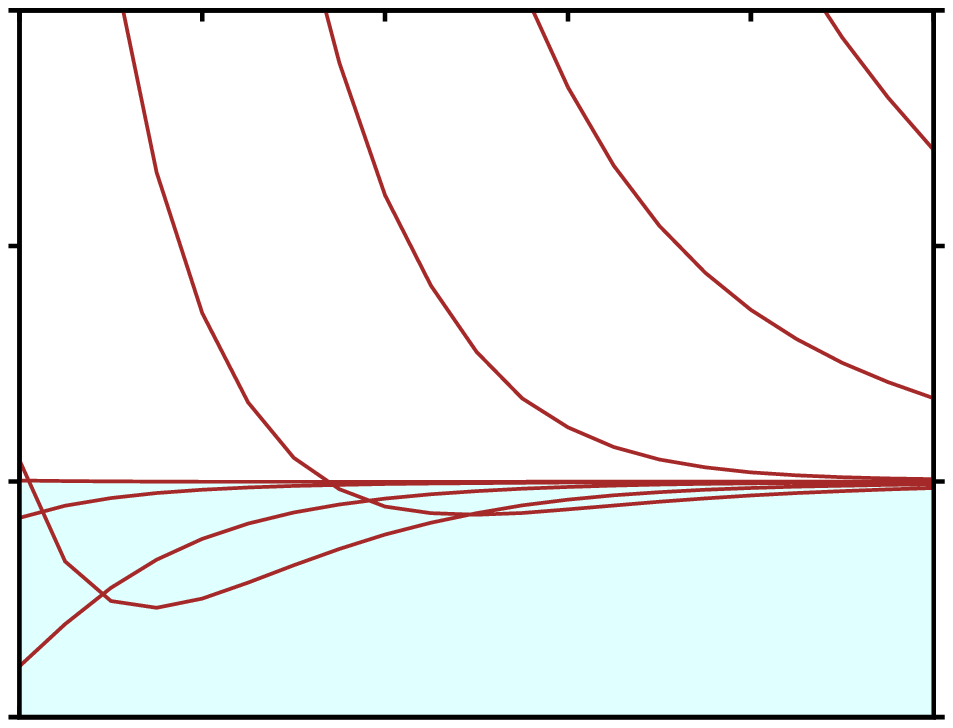} } 
 \end{array}$
 \caption{Modal diagram for HD\,98851, where $T_{\rm eff}$ = 7000 K and $\log L/L_{\odot}$ = 1.43.}
 \normalsize
 \label{m_hd98851}
 \end{figure*}

\begin{figure*}
\centering $
\large
\begin{array}{cc}
   \scalebox{0.66}{ \input{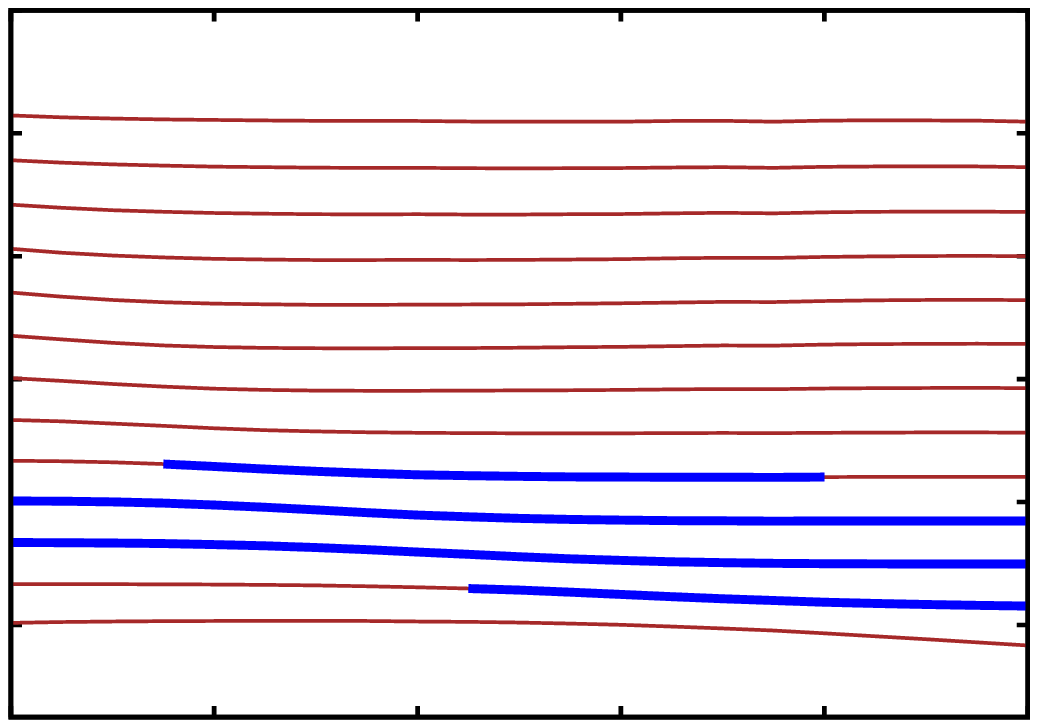} } 
    \scalebox{0.66}{ \input{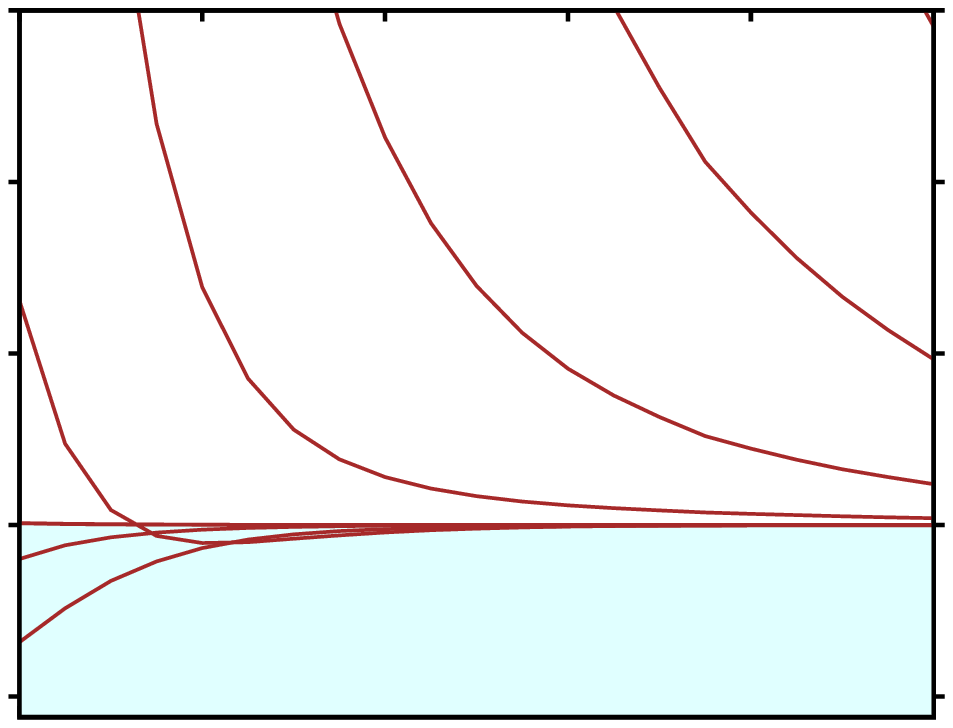} } 
 \end{array}$
 \caption{Modal diagram for HD\,118660, where $T_{\rm eff}$ = 7550 K and $\log L/L_{\odot}$ = 1.12.}
 \normalsize
 \label{m_hd118660}
 \end{figure*}

\begin{figure*}
\centering $
\large
\begin{array}{cc}

   \scalebox{0.66}{ \input{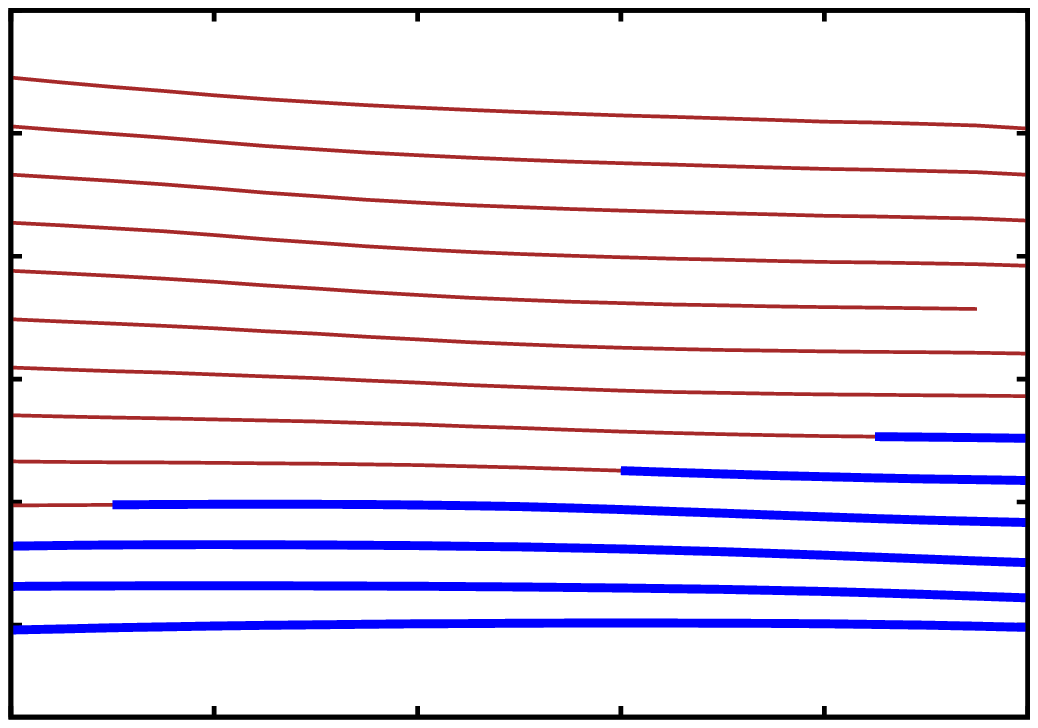} } 
    \scalebox{0.66}{ \input{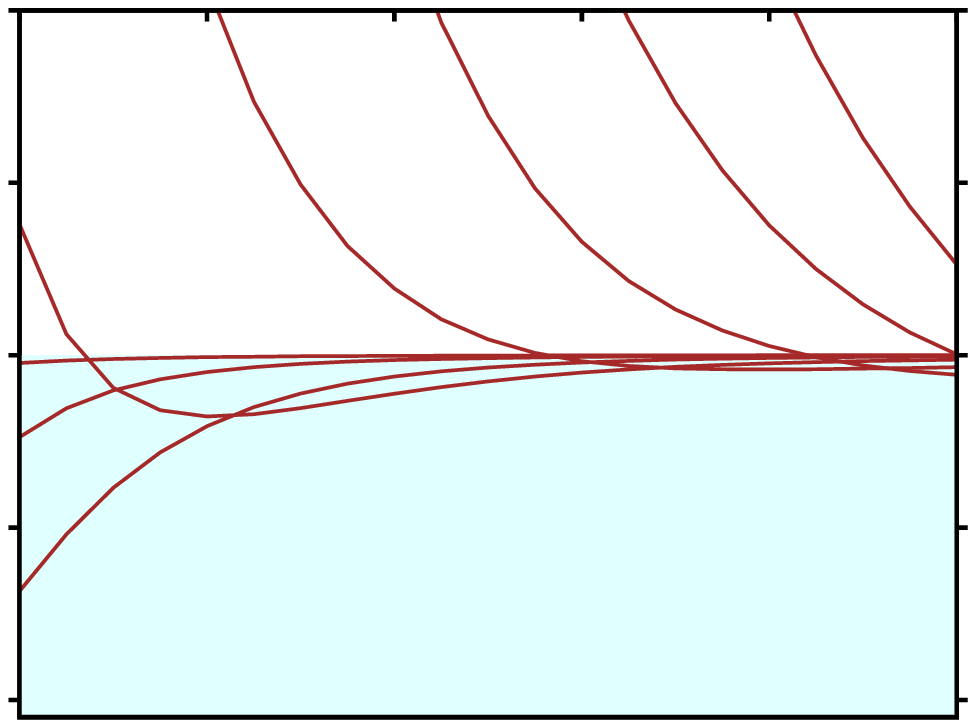} } 
 \end{array}$
 \caption{Modal diagram for HD\,102480, where $T_{\rm eff}$ = 6720 K and $\log L/L_{\odot}$ = 1.41.}
 \normalsize
 \label{m_hd102480}
 \end{figure*}

\begin{figure*}
\centering $
\large
\begin{array}{cc}

   \scalebox{0.66}{ \input{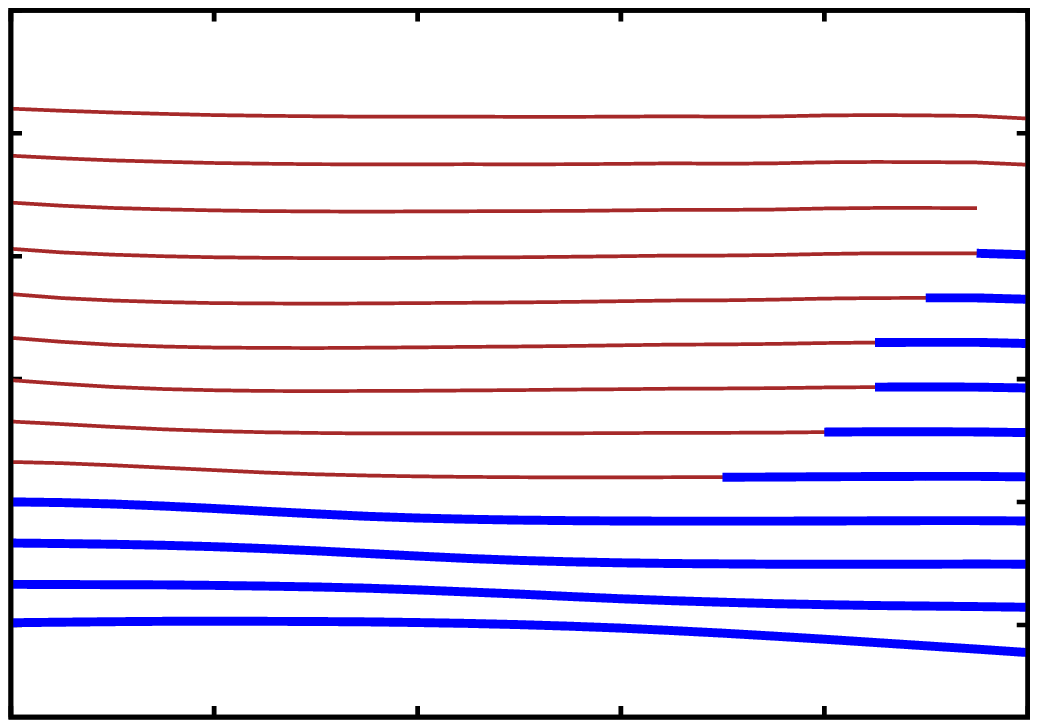} } 
    \scalebox{0.66}{ \input{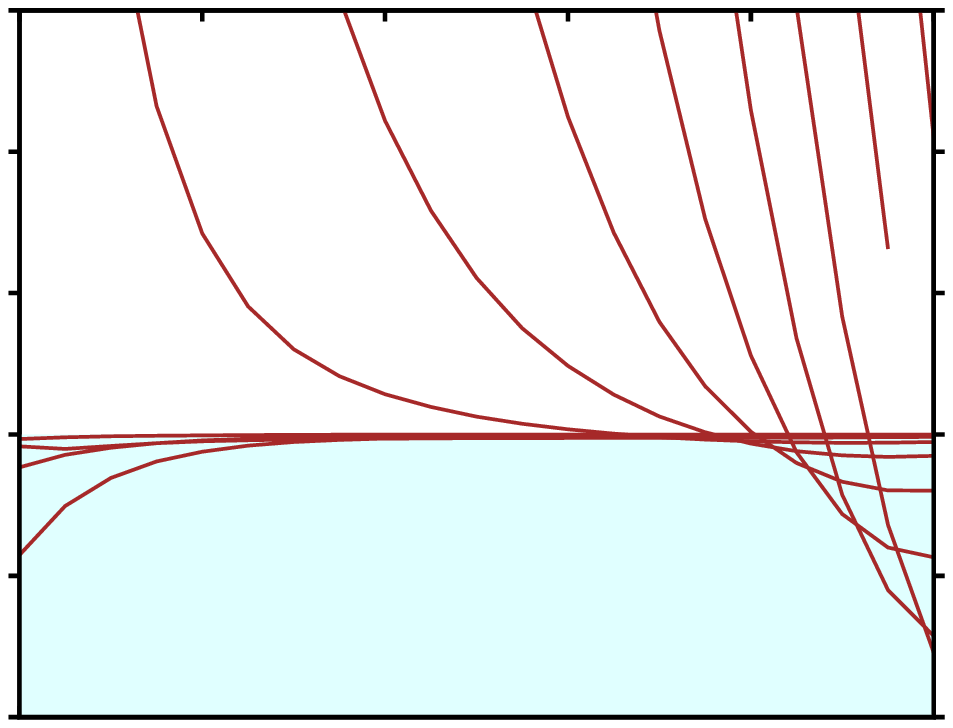} } 
 \end{array}$
 \caption{Modal diagram for HD\,13079, where $T_{\rm eff}$ = 7040 K and $\log L/L_{\odot}$ = 0.97.}
 \normalsize
 \label{m_hd13079}
 \end{figure*}

\begin{figure*}
\centering $
\large
\begin{array}{cc}

   \scalebox{0.66}{ \input{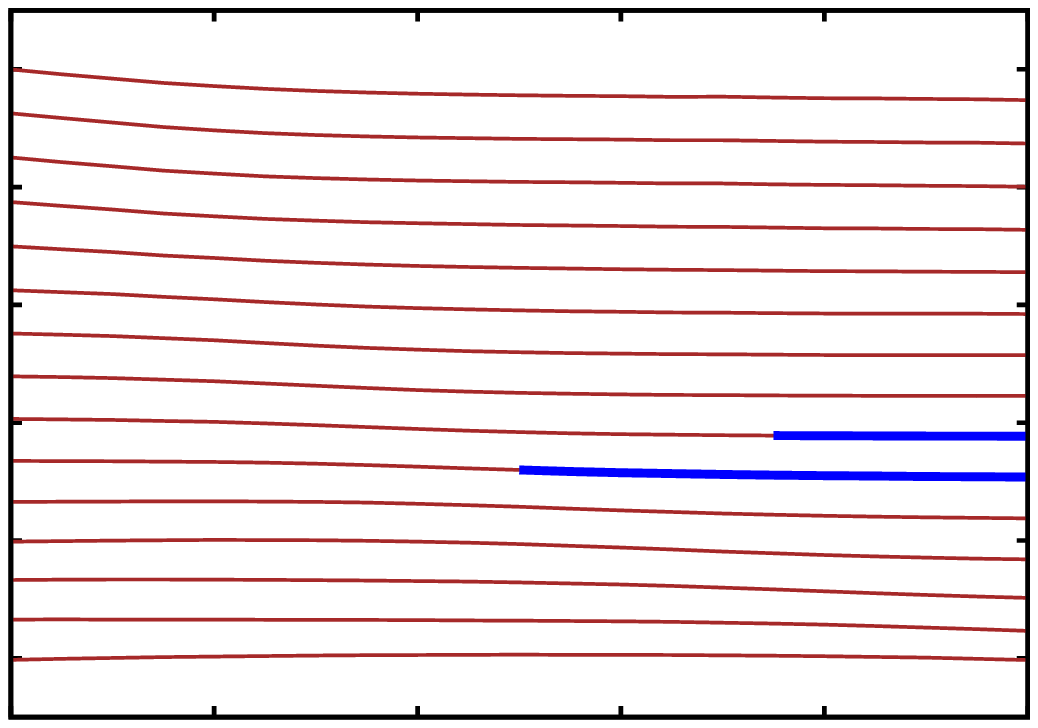} } 
    \scalebox{0.66}{ \input{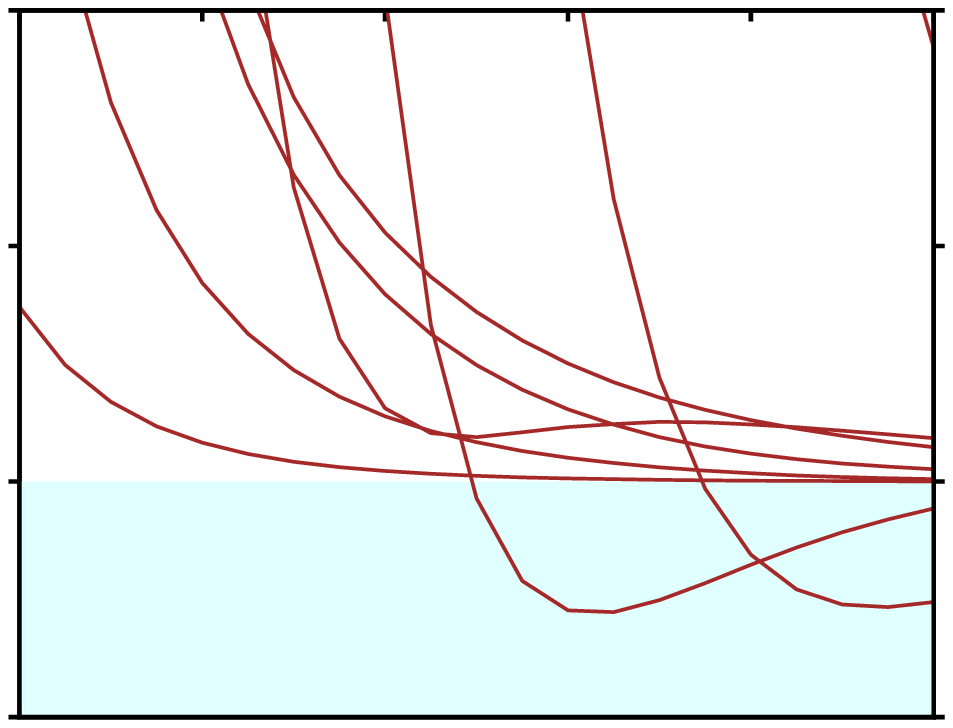} } 
 \end{array}$
 \caption{Modal diagram for HD\,13038, where $T_{\rm eff}$ = 7960 K and $\log L/L_{\odot}$ = 1.51.}
 \normalsize
 \label{m_hd13038}
 \end{figure*}

\begin{figure*}
\centering $
\large
\begin{array}{cc}

   \scalebox{0.66}{ \input{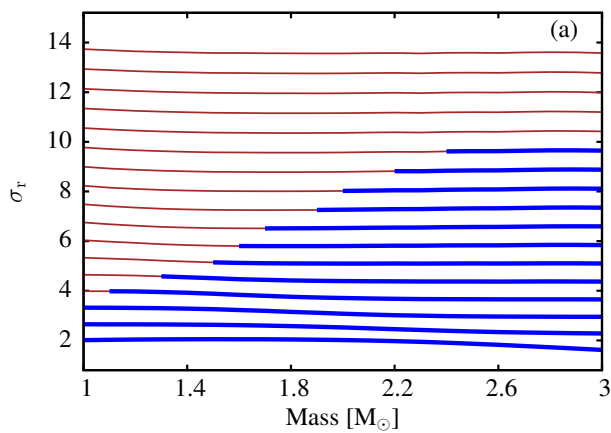} } 
    \scalebox{0.66}{ \input{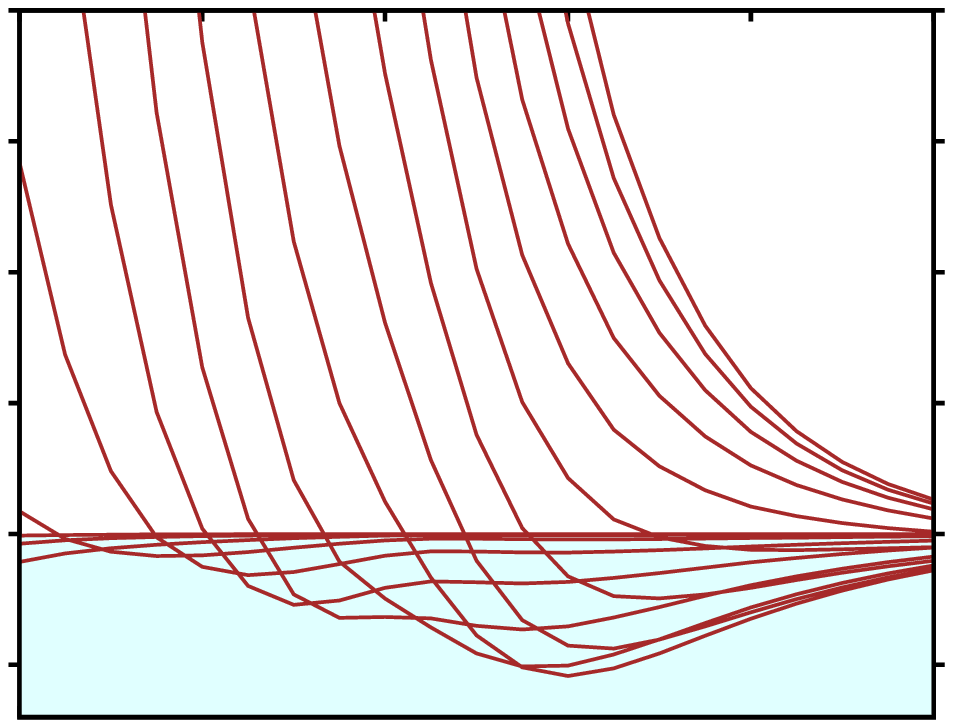} } 
 \end{array}$
 \caption{Modal diagram for HD\,25515, where $T_{\rm eff}$ = 6650 K and $\log L/L_{\odot}$ = 1.01.}
 \normalsize
 \label{m_hd25515}
 \end{figure*}

Similar to HD\,113878, results of the linear stability analysis for the models of HD\,98851,
HD\,118660, HD\,102480, HD\,13079, HD\,13038, and HD\,25515 are given in the Figs.\,\ref{m_hd98851}, \ref{m_hd118660}, \ref{m_hd102480}, \ref{m_hd13079},
\ref{m_hd13038}, and \ref{m_hd25515}, respectively.
Except for the star HD\,13038, low order p-modes are unstable in all models within the considered mass-range of these stars. For HD\,13038,
unstable modes are only found for models with $M > 2$\,M$_{\odot}$ (Fig.\,\ref{m_hd13038}). 
HD\,25515 is the star with the largest number of unstable modes (Fig.\,\ref{m_hd25515}).

 
 \begin{figure*}
\centering $
\large
\begin{array}{cc}

   \scalebox{0.66}{ \input{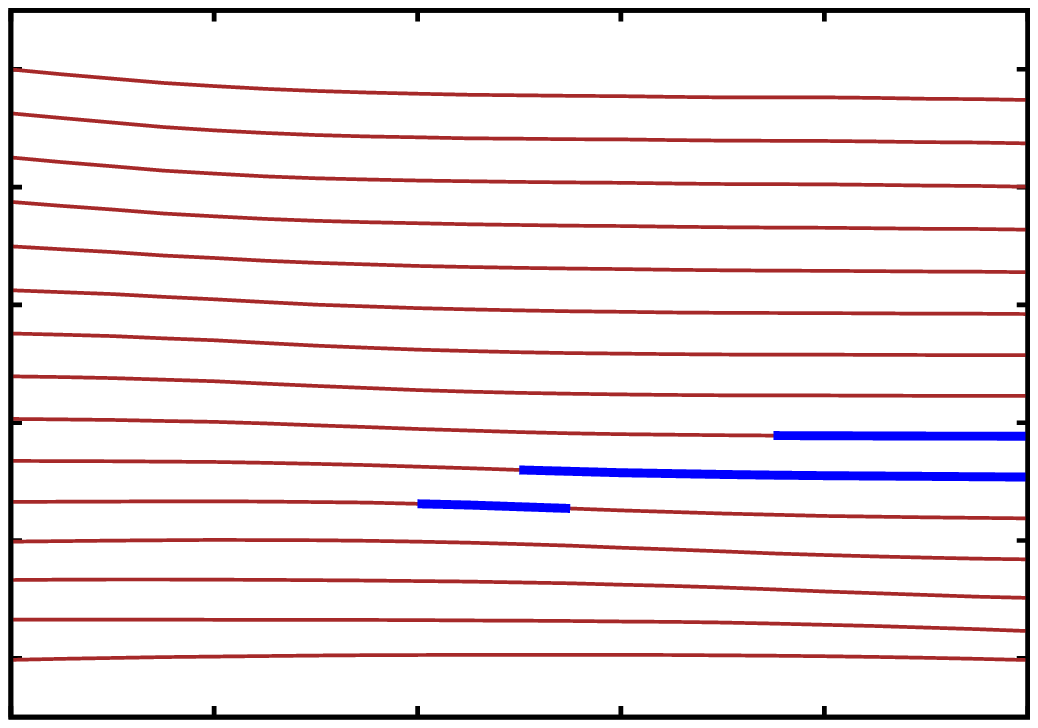} } 
    \scalebox{0.66}{ \input{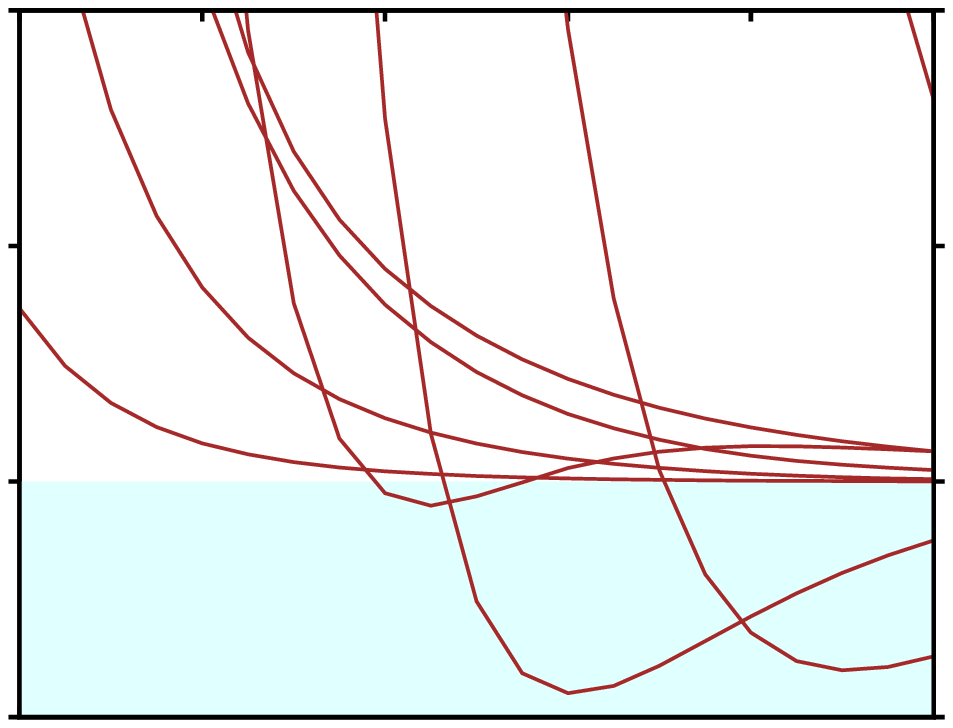} } 
 \end{array}$
 \caption{Same as Fig.\,\ref{m_hd13038} but for mixing length parameter $\alpha$ = 1.8 (for HD\,13038).}
 \normalsize
 \label{alfa18}
 \end{figure*}

 \begin{figure*}
\centering $
\large
\begin{array}{cc}

   \scalebox{0.66}{ \input{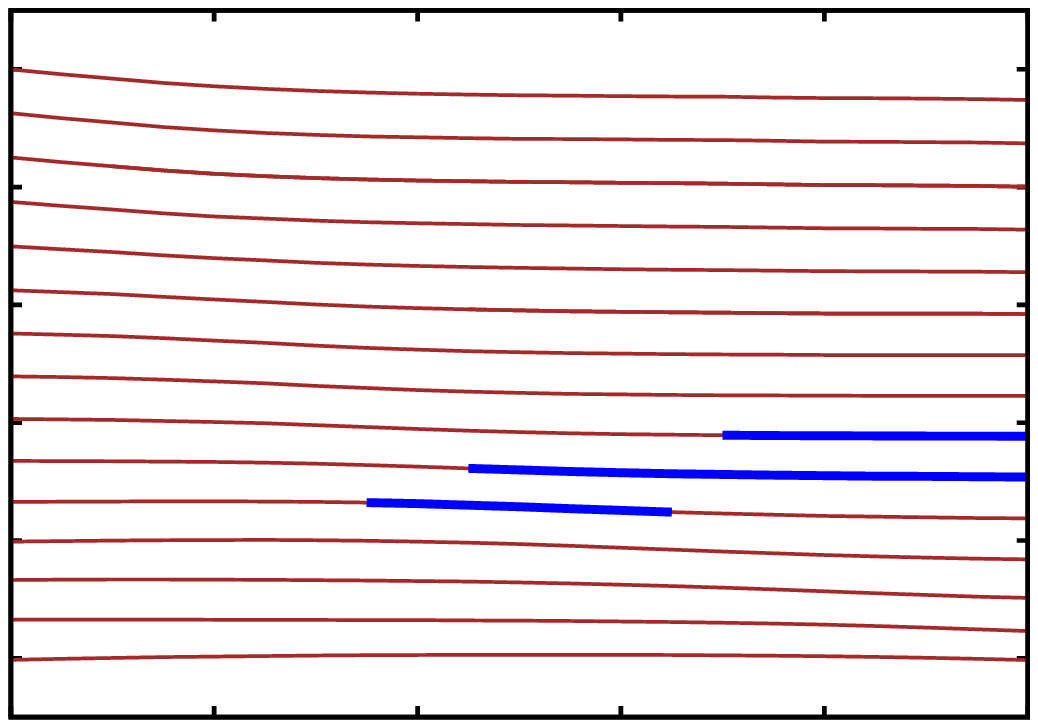} } 
    \scalebox{0.66}{ \input{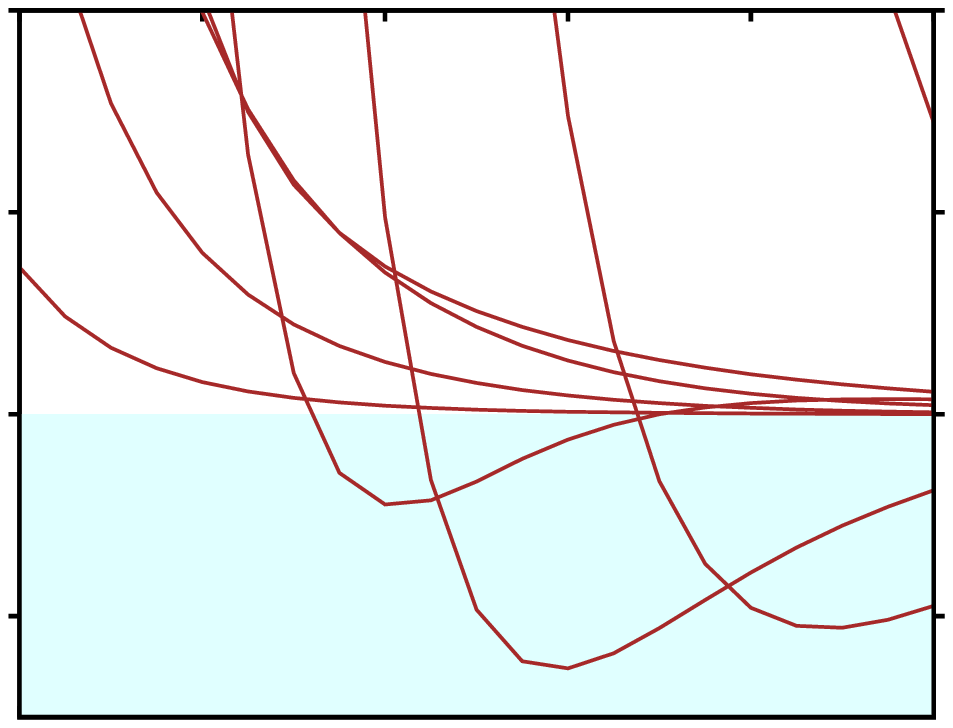} } 
 \end{array}$
 \caption{Same as Fig.\,\ref{m_hd13038} but for mixing length parameter $\alpha$ = 2.0 (for HD\,13038).}
 \normalsize
 \label{alfa20}
 \end{figure*}

 Considering the perturbation of convective flux, Dupret et al. (2004, 2005) have obtained theoretical instability strips for 
 $\delta$\,Sct stars. These authors have pointed out that the theoretical instability strip is sensitive to the opted value of $\alpha$.
 So far, we have used $\alpha$\,=\,1.5 in the present study. 
 In order to examine the effect of  
 $\alpha$ on the instabilities, we have calculated two additional sets of models for the star HD\,13038. 
 Results of the linear stability analysis for models with $\alpha$ = 1.8 and 2.0 are given in Fig.\,\ref{alfa18} and Fig.\,\ref{alfa20},
 respectively. 
 From a comparison with Fig.\,\ref{m_hd13038}, we learn 
 that for models with $\alpha$ = 1.8 
 and 2.0, one additional mode is unstable and the instability region is extended to the lower mass models, 
 confirming the sensitivity of the results to the value of $\alpha$.

\begin{figure*}
\centering $
\large
\begin{array}{c}

   \scalebox{0.8}{ \input{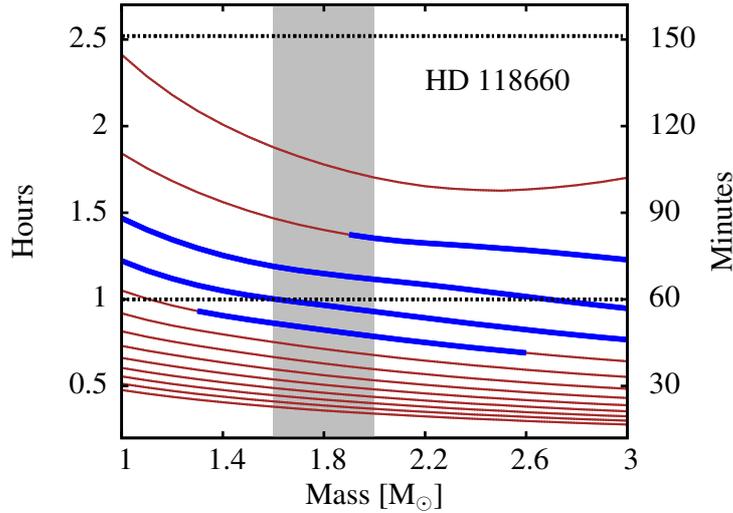} } 
 \end{array}$
 \caption{Periods associated with different modes as a function of stellar mass for the models of HD\,118660. Thick blue lines correspond 
 to the unstable modes and dotted lines represent the observed periods. The gray column represents a range of mass associated with
 the evolutionary tracks which are compatible with the observed error bar in the HRD for this star.}
 \normalsize
 \label{periods_118660}
 \end{figure*}

 In all these seven stars, the observed periods are in the range of several minutes to a few hours (Joshi et al. 2017).
The pulsation periods associated with the unstable radial modes found in models of these 
stars are also in the same range. Fig.\,\ref{periods_118660} shows the pulsation periods corresponding to the different low order
radial modes as a function of stellar mass for HD\,118660. Thick 
blue lines represent unstable modes while the observed pulsation periods of this star 
are displayed with dotted lines. A gray column depicts a range of mass (associated with the evolutionary tracks) which is compatible with 
the observed error box for HD\,118660 in the HRD (Fig.\,\ref{hrd}).
The observed period of 60 minutes is consistent with an unstable radial mode for models with a mass 
close to 1.6 and 2.7 M$_{\odot}$ (Fig.\,\ref{periods_118660}).
However, the observed period of 151.2 minutes can not be explained with unstable radial modes
and probably corresponds to an unstable non-radial pressure mode.
The location of HD\,118660 in the HRD (Fig.\,\ref{hrd}) suggests that the mass of this star is smaller than 2 M$_{\odot}$, 
from which we conclude that the mass of HD\,118660 is close to 1.6\,M$_{\odot}$ if a radial 
mode is responsible for the observed 60-min period.
However, a linear stability analysis for 
non-radial perturbations is required to
suffice this conclusion.

\begin{figure*}
\centering $
\large
\begin{array}{cccccc}
   \scalebox{0.66}{ \input{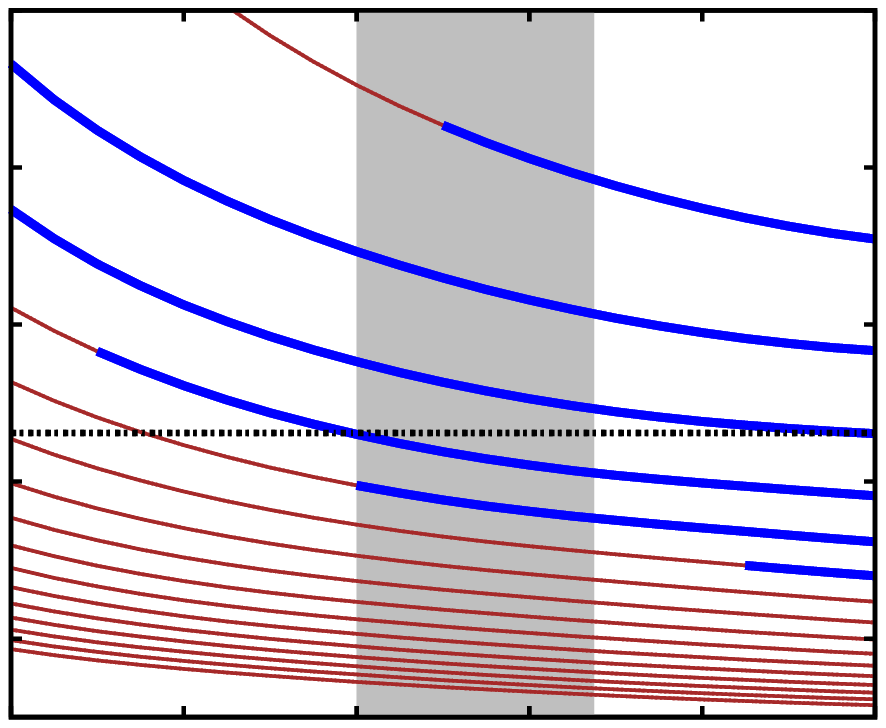} } 
   \scalebox{0.66}{ \input{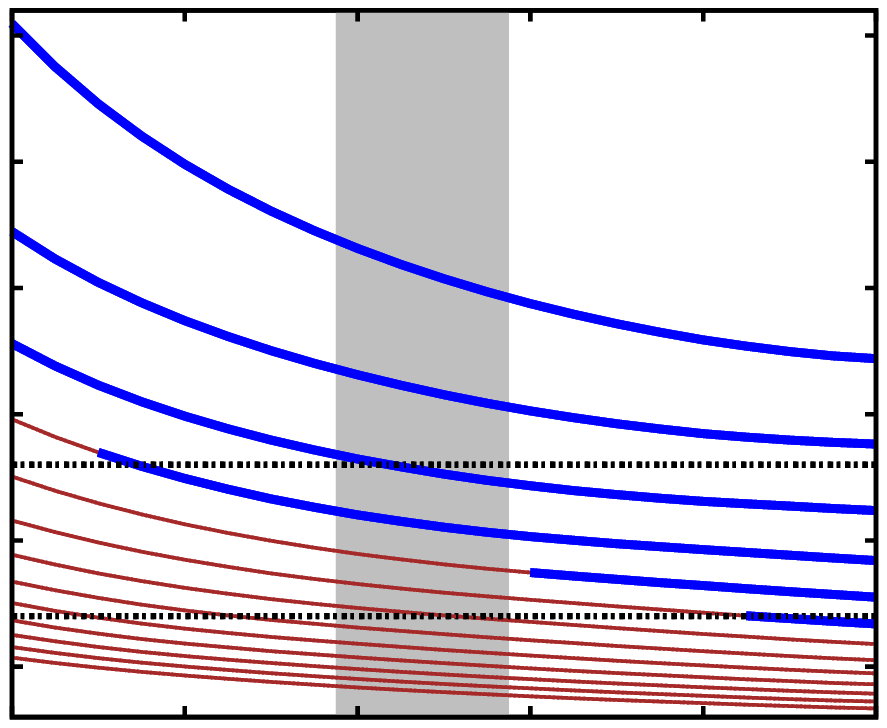} } \\
   \scalebox{0.66}{ \input{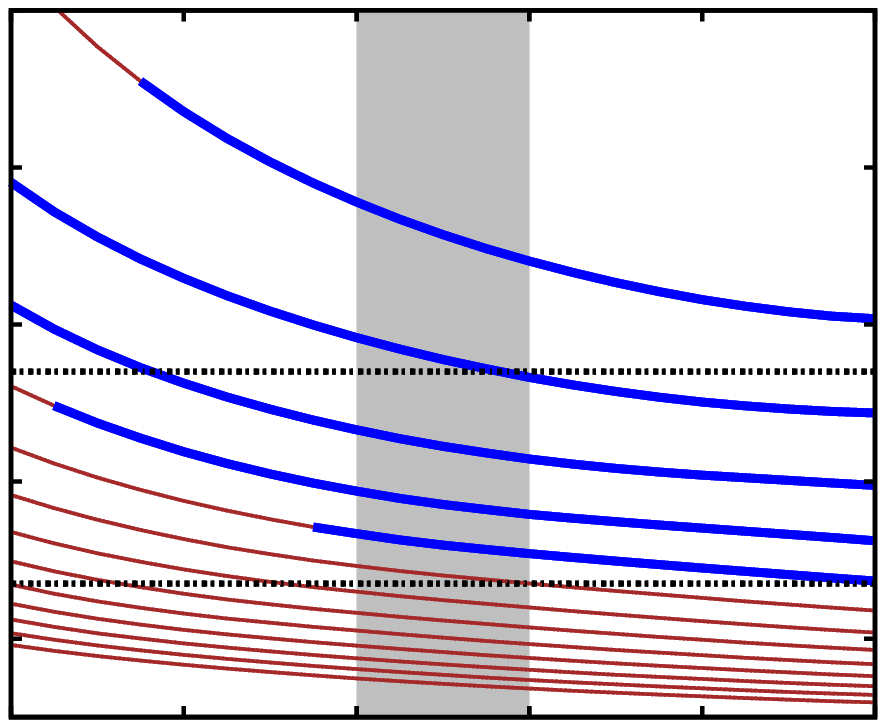} } 
   \scalebox{0.66}{ \input{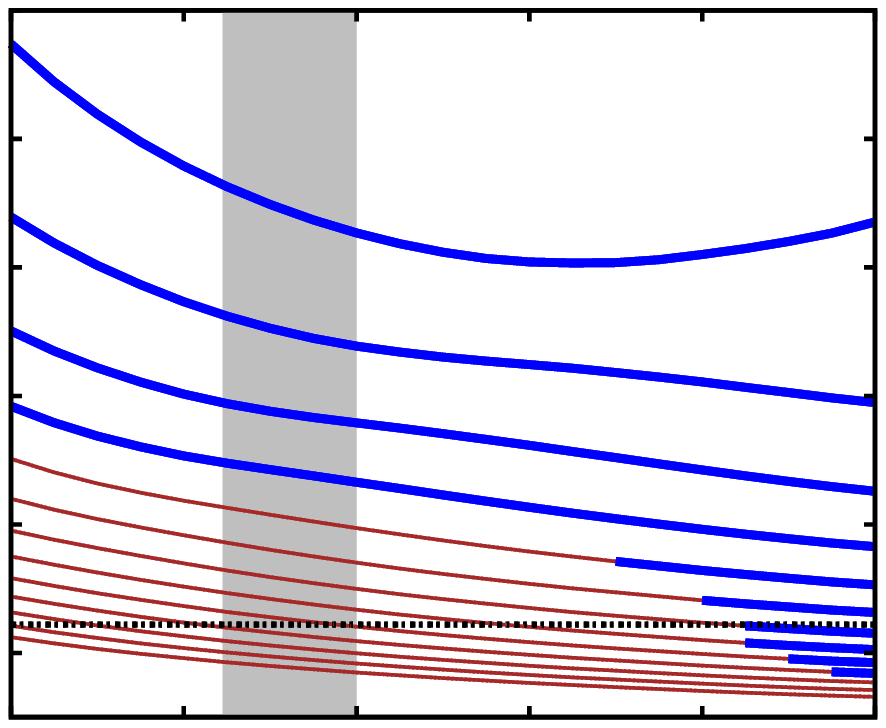} } \\
   \scalebox{0.66}{ \input{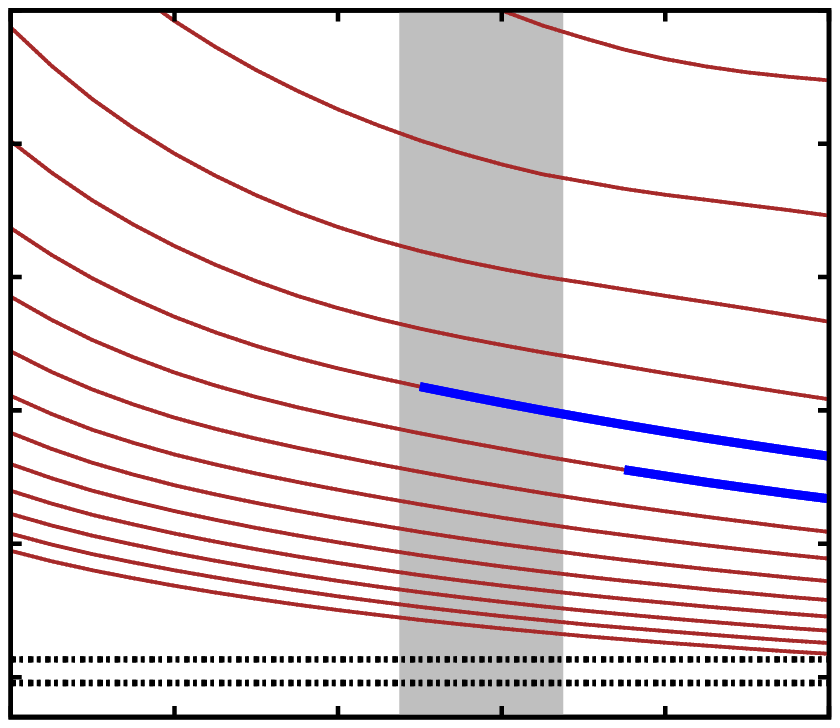} } 
   \scalebox{0.66}{ \input{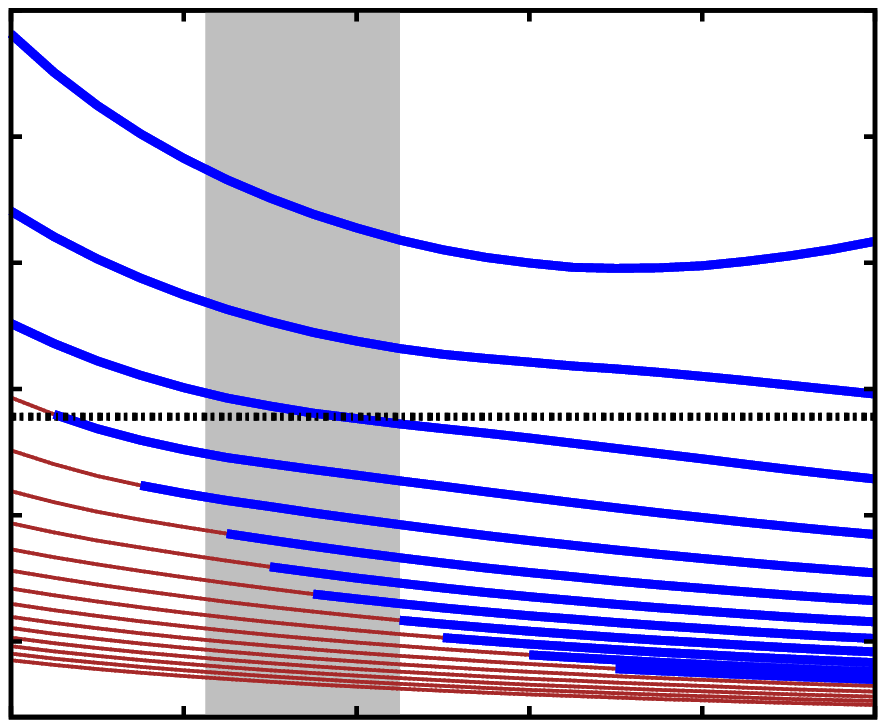} } \\
 \end{array}$
 \caption{Periods associated with different modes as a function of stellar mass for the models of
HD\,113878, HD\,102480, HD\,98851, HD\,13079, HD\,13038 and HD\,25515. 
Thick blue lines correspond to the unstable modes and dotted lines represent the observed
periods. Gray column represents the range of mass associated with
 the evolutionary tracks which are compatible with the observed error bar in the HRD for the considered stars.
}
 \normalsize
 \label{periods}
 \end{figure*}

Similar to Fig.\,\ref{periods_118660} for HD\,118660, pulsation periods associated with the low order radial modes for the stars HD\,113878, HD\,102480, HD\,98851,
HD\,13079, HD\,13038, and HD\,25515 are given in Fig.\,\ref{periods}. For HD\,113878, a variability of 
138.6 minutes has been observed that 
can be explained by the low order radial mode excited in models of HD\,113878 having mass of $\sim$ 1.8 M$_{\odot}$ (Fig.\,\ref{periods}, top left). 
The observed period of 156 minutes
in the photometric data 
of HD\,102480 can be elucidated with the help of an unstable radial mode found in models with a mass near 1.85 M$_{\odot}$
(Fig.\,\ref{periods}, top right). 
An unstable radial mode consistent with the period of 156 minutes is also present in models of HD\,102480 having mass 
close to 1.3 M$_{\odot}$. In 
addition to the period of 156 minutes, another period of 84 minutes has been observed for HD\,102480 which is consistent with an unstable 
radial mode present 
in models of this star with a mass near to 2.7 M$_{\odot}$. However, considering the compatible mass range (1.75 M$_{\odot}$ to 
2.15 M$_{\odot}$) represented by the gray region 
in Fig.\,\ref{periods} (top right), we can 
exclude the unstable modes present in models with a mass close to 1.3 M$_{\odot}$ and 2.7 M$_{\odot}$ as an explanation for the 
observed variabilities in HD\,102480.
A model of 
HD\,98851 with a mass of 2.1 M$_{\odot}$ has an unstable radial mode whose period is exactly matching with the observed period of 162 minutes 
(Fig.\,\ref{periods}, middle left). An observed period of 81 minutes for HD\,98851 is consistent with an unstable 
radial mode present in models having mass 
$\sim$ 2.9 M$_{\odot}$ but this solution can be disregarded as the mass of these models is far from the compatible mass range 
(1.8 M$_{\odot}$ to 2.2 M$_{\odot}$)
found for this star (Fig.\,\ref{periods}, middle left). Variability
with a period of 73.2 minutes has been reported for HD\,13079 (Fig.\,\ref{periods}, middle right). In our 
stability analysis, we do not find any 
unstable radial
mode capable to explain this observed period for
models having mass in the compatible mass range (1.49 M$_{\odot}$ to 1.80 M$_{\odot}$) of  
HD\,13079. Only models with a mass greater than 2.7 M$_{\odot}$ are consistent with the observed period. 
In the case of HD\,13038, two periods of 28.7 and 34.0 minutes have been observed (Fig.\,\ref{periods}, bottom left). In the 
stability analysis, 
we have not found unstable radial modes in this period range in the models of this star. 
Variability with a period of 166.8 minutes has been reported in the data of HD\,25515 
which can be interpreted by an unstable radial mode found in models of the star with a mass close to 1.8 M$_{\odot}$ 
(Fig.\,\ref{periods}, bottom right). For this star, models having mass close to 1.1 M$_{\odot}$ also have an unstable mode 
with a period of 166.8 minutes but this mass is outside the compatible mass range (1.45 M$_{\odot}$ to 1.90 M$_{\odot}$) of 
HD\,25515. This solution can be excluded for further consideration.  
We therefore conclude that if the observed period of 166.8 minutes is due to a radial mode, the mass of HD\,25515 is
close to 1.8 M$_{\odot}$.

\section{Discussion and conclusions}
\label{d}
In the present study, we have attempted the first step to understand the pulsational variabilities observed under the Nainital-Cape
Survey for low and intermediate mass stars with properties that place them in the theoretical instability strip of $\delta$\,Sct stars
(Fig.\,\ref{hrd}). 
We have constructed 
the envelope models of 7 stars and performed a non-adiabatic linear stability analysis for these models. For the construction of the stellar models,  
we have used the values of the required input parameters published by Joshi et al. (2017). 
Results of the linear stability analysis have been 
discussed in terms of modal diagrams where eigenfrequencies are given as a function of stellar mass. 
Generally, modes with higher eigenfrequencies correspond to the higher order modes. 
From the modal diagrams (Figs.\,\ref{m_hd113878} - \ref{m_hd25515}), we deduce that several low order radial modes 
are unstable while the high order
modes are mostly damped.
The periods associated with the unstable modes 
are in the range of several minutes to a few hours which is in agreement with the observed periods of the $\delta$\,Sct pulsators. For 
example, with the help of photometry, Joshi et al. (2006) have reported a pulsation period of 60 minutes in HD\,118660. In our study, 
the unstable modes found in models of HD\,118660 have pulsation periods equivalent to this observed value 
(Fig.\,\ref{periods_118660}). Similarly for the stars HD\,113878, HD\,102480, HD\,98851, and HD\,25515, most of the observed variabilities are 
found to have periods which can be explained using the low order radial modes in models with stellar parameters consistent with 
the observed ones for these stars (Fig.\,\ref{periods}).

In the frequency spectra of the evolutionary stellar models with a mass of 1.8 M$_{\odot}$, Pamyatnykh (2000) and Dupret et al. (2005) have 
reported the presence of low order unstable radial modes where consecutive modes are separated with almost an equivalent 
distance in frequency (e.g., Fig.\,1 of Pamyatnykh 2000 and Dupret et al. 2005). The outcome of the present study is in 
accordance with these previous 
findings (Figs.\,\ref{m_hd113878} - \ref{m_hd25515}). Dupret et al. (2004) have pointed out that the value of mixing length parameter $\alpha$ 
can influence the instability and location of the instability strip of $\delta$\,Sct stars. All our calculations
have been done for $\alpha$\,=\,1.5 pressure
scale heights, but for HD\,13038, we have also performed additional linear stability analyses
for $\alpha$\,=\,1.8 and 2.0. For these extra values of $\alpha$, one additional mode is found to be 
unstable and the strength of the associated instability increases with the value of  $\alpha$ (Figs.\,\ref{alfa18} and \ref{alfa20}).

The present study is restricted to radial perturbations. As non-radial pulsations have also been found in 
several $\delta$\,Sct stars, an extensive linear stability analysis with non-radial perturbations will be very useful to
understand the observed 
variabilities in the considered stars. Apart from that, the effect of chemical compositions on the instabilities will also be of 
particular interest.

The Nainital-Cape survey has been a bilateral scientific effort between India and South Africa. So far, observational studies of Ap and Am
type stars have been the primary aim of this project.  In order to understand the reported variabilities (as well as null results) 
in the observed stars, asteroseismic modelling is required and this study is a first step in this direction. Observing facilities as 
well as the expertise related to asteroseismic modelling within the Belgo-Indian Network for Astronomy and astrophysics (BINA) will be
very useful to extend the studies done under the Nainital-Cape survey. Moreover, a continuation of this project with the involvement
of the BINA consortium will not only boost the quality of scientific outputs but also foster a tri-nation (Belgium, India and South Africa) 
collaboration.

% USE A SECTION WITHOUT NUMBER FOR THE ACKNOWLEDGEMENTS
%
\section*{Acknowledgements}
The author thanks Dr. Santosh Joshi and Dr. Peter De Cat for their continuous encouragement and support. The author is grateful to 
Prof. Wolfgang Glatzel for earlier discussions on the topic 
presented in this paper. 
Constructive comments and suggestions of the referee are gratefully acknowledged.   
%
% BEGIN THE REFERENCE LIST WITH \beginrefer
% USE \refer BEFORE THE REFERENCES AND BEGIN A NEW PARAGRAPH AFTER THE 
% REFERENCE !
% DO NOT FORGET TO END THE LIST WITH \endrefer
% 
%
% INSTRUCTIONS FOR BIBLIOGRAPHY:
% ==============================
% - DON'T USE THE & SYMBOL
% - USE INITIALS FOR FIRST AND MIDDLE NAMES, AND SPECIFY FULL FAMILY NAME (see examples below)
% - NO COMMA BETWEEN NAME AND INITIALS
% - USE COMMA BETWEEN DIFFERENT AUTHORS NAMES
% - NO COMMA AFTER THE LAST AUTHOR NAME
% - FOR LONG AUTHOR LISTS, SPECIFY THE FIRST 3 AUTHORS FOLLOWED BY 'et al.', WITH NO COMMA BEFORE AND AFTER 'et al.'
% - INSERT A BLANK SPACE BETWEEN MULTIPLE INITIALS
% - USE STANDARD JOURNAL ACRONYMS FREQUENTLY USED IN MAIN ASTROPHYSICS JOURNAL
% - SORT REFERENCES BY ALPHABETICAL ORDER OF FIRST AUTHOR NAMES

\footnotesize
%\bibliographystyle{aa}
%\bibliography{first}

\beginrefer

\refer Aerts C., Christensen-Dalsgaard J., Kurtz D. W. 2010, Asteroseismology, Astronomy and Astrophysics Library. Springer, Berlin Heidelberg

\refer Ashoka B. N., Seetha S., Raj E. et al. 2000, BASI, 28, 251

\refer Balona L. A., Catanzaro G., Crause L. et al. 2013, MNRAS, 432, 2808

\refer Balona L. A., Engelbrecht C. A., Joshi Y. C. et al. 2016, MNRAS, 460, 1318

\refer Baker N., Kippenhahn R. 1965, ApJ, 142, 868

\refer B\"ohm-Vitense E. 1958, ZA, 46, 108

\refer Breger, M., Lenz, P., Antoci, V., et al. 2005, A\&A, 435, 955

\refer Dupret M., Grigahc\`ene A., Garrido R., Gabriel M., Scuflaire R., 2004, A\&A, 414, L17

\refer Dupret M., Grigahc\`ene A., Garrido R., Gabriel M., Scuflaire R., 2005, A\&A, 435, 927

\refer Gautschy A., Glatzel W. 1990a, MNRAS, 245, 154

\refer Gautschy A., Glatzel W. 1990b, MNRAS, 245, 597

\refer Glatzel W., Kiriakidis M. 1993, MNRAS, 262, 85

\refer Glatzel W., Kiriakidis M., Fricke K. J. 1993, MNRAS, 262, L7

\refer Iglesias C. A., Rogers F. J. 1996, ApJ, 464, 943

\refer Kippenhahn R., Weigert A., Weiss A. 2012, Stellar Structure and Evolution, Astronomy and Astrophysics Library. Springer, Berlin Heidelberg

\refer Kiriakidis M., Fricke K. J., Glatzel W. 1993, MNRAS, 264, 50

\refer Joshi S., Girish V., Sagar R. et al. 2003, MNRAS, 344, 431

\refer Joshi S., Mary D. L., Martinez P. et al. 2006, A\&A, 455, 303

\refer Joshi S., Mary D. L., Chakradhari N. K. et al. 2009, A\&A, 507, 1763

\refer Joshi S., Ryabchikova T., Kochukhov O. et al. 2010, MNRAS, 401, 1299

\refer Joshi S., Semenko E., Martinez P. et al. 2012a, MNRAS, 424, 2002

\refer Joshi S., Martinez P., Chowdhury S. et al. 2016, A\&A, 590, A116

\refer Joshi S., Semenko E., Moiseeva A. et al. 2017, MNRAS, 467, 633

\refer Joshi Y. C., Joshi S., Kumar B. et al. 2012b, MNRAS, 419, 2379

\refer Joshi Y. C., Balona L. A., Joshi S. et al. 2014, MNRAS, 437, 804

\refer Martinez P., Kurtz D. W., Ashoka B. N. et al. 2001, A\&A, 371, 1048

\refer Pamyatnykh, A. A. 2000, ASP Conf. Ser., 210, 215

\refer Rogers F. J., Iglesias C. A. 1992, ApJS, 79, 507

\refer Rogers F. J., Swenson F. J., Iglesias C. A. 1996, ApJ, 456, 902

\refer Saio H., Baker N. H., Gautschy A. 1998, MNRAS, 294, 622

\refer Yadav A. P., Glatzel W. 2017a, MNRAS, 465, 234

\refer Yadav A. P., Glatzel W. 2017b, MNRAS, 471, 3245

\endrefer

\end{document}